\documentclass[10pt,journal]{IEEEtran}

\usepackage[hyphens]{url}  % DO NOT CHANGE THIS
\usepackage{graphicx} % DO NOT CHANGE THIS
\usepackage{newfloat}
\usepackage{listings}
\usepackage{amsmath,amstext,amsfonts,stmaryrd,bm}
\usepackage{mathtools}
\usepackage{color}

\newcommand{\Time}{\ensuremath{\mathbb{T}}}

\newcommand{\ZPos}{\ensuremath{\mathbb{Z}^{+}_{0}}}

\newcommand{\Meets}[2]{\ensuremath{#1 \sim #2}}
\newcommand{\MetBy}[2]{\ensuremath{#1 \sim^{\star} #2}}
\newcommand{\Before}[2]{\ensuremath{#1 \prec #2}}
\newcommand{\After}[2]{\ensuremath{#1 \succ #2}}
\newcommand{\Contains}[2]{\ensuremath{#1 \supset #2}}
\newcommand{\During}[2]{\ensuremath{#1 \subset #2}}
\newcommand{\StartedBy}[2]{\ensuremath{#1 \uparrow^{\star} #2}}
\newcommand{\Starts}[2]{\ensuremath{#1 \uparrow #2}}
\newcommand{\FinishedBy}[2]{\ensuremath{#1 \downarrow^{\star} #2}}
\newcommand{\Finishes}[2]{\ensuremath{#1 \downarrow #2}}
\newcommand{\OverlappedBy}[2]{\ensuremath{#1 \frown^{\star} #2}}
\newcommand{\Overlaps}[2]{\ensuremath{#1 \frown #2}}
\newcommand{\Disjoint}[2]{\ensuremath{#1 \mid\mid #2}}
\newcommand{\In}[2]{\ensuremath{#1 \sqsupset #2}}

\newcommand{\Atoms}[1]{\ensuremath{\mathrm{Atoms}(#1)}}

\newcommand{\Int}{\mathsf{Int}}
\newcommand{\IntRel}[2]{\ensuremath{\mathsf{#1\,\mathsf{Int}\,#2}}}

\newcommand{\ipstarts}[3]{\ensuremath{\mathrm{starts}(#1, #2, #3)}}
\newcommand{\ipends}[3]{\ensuremath{\mathrm{ends}(#1, #2, #3)}}
\newcommand{\ipspans}[4]{\ensuremath{\mathrm{spans}(#1, #2, #3, #4)}}
\newcommand{\ipcontains}[3]{\ensuremath{\mathrm{contains}(#1, #2, #3)}}

\newtheorem{axiom}{Axiom}
\newtheorem{theorem}{Theorem}

\newtheorem{definition}{Definition}

\newtheorem{remark}{Remark}
\newtheorem{assumption}{Assumption}

\begin{document}

\title{Temporal Planning via Interval Logic Satisfiability for Autonomous Systems}

\author{Miquel~Ramirez, Anubhav~Singh, Peter~J.~Stuckey, Chris~Manzie
\thanks{Manuscript uploaded to Arxiv on 14th June 2024 by M. Ramirez.}
\thanks{M. Ramirez is with The University of Melbourne, Dept. of Electrical and Electronic Engineering, Parkville VIC 3052 Australia (e-mail: firstname.lastname@unimelb.edu.au)  }
\thanks{A. Singh is with The University of Toronto, Dept. of Mechanical and Industrial Engineering, Toronto ON M5S 3G8 Canada}
\thanks{P. J. Stuckey is with Monash University, Dept. of Data Science \& AI, Clayton VIC 3168 Australia}
\thanks{C. Manzie is with The University of Melbourne, Dept. of Electrical and Electronic Engineering, Parkville VIC 3052 Australia}
}
% For research notes, remove the comment character in the line below.
% \researchnote

\maketitle

\begin{abstract}
Many automated planning methods and formulations rely on suitably designed abstractions or simplifications of the
constrained dynamics associated with agents to attain computational scalability. We consider formulations of
temporal planning where intervals are
associated with both action and fluent atoms, and relations between these are given as sentences in Allen’s Interval
Logic. We propose a notion of planning graphs that can account for complex concurrency relations between actions and
fluents as a Constraint Programming (CP) model. We test an implementation of our algorithm
on a state-of-the-art framework for CP and compare it with PDDL 2.1 planners that capture plans requiring complex
concurrent interactions between agents. We demonstrate our algorithm outperforms existing PDDL 2.1 planners in the
case studies. Still, scalability remains challenging when plans must comply with intricate concurrent interactions
and the sequencing of actions.
\end{abstract}

\section{Introduction}
\label{Intro}
Temporal planning is an optimization problem where solutions are given by sets of actions, or \emph{plans},
which need to be chosen in such a way that certain requirements are met.
Actions abstract the set of possible behaviours that are available to steer a given dynamical system.
We distinguish three types of requirements on actions.
The first is that of \emph{suitability and relevance}
of an action to bring about some task or goal.
For instance, we may need to transport cargo between locations, using as efficiently as possible a given
set of automated vehicles.
Actions that move vehicles between locations are relevant as long as a vehicle has been tasked with
running cargo between those locations. Other actions that abstract the handling of cargo become relevant
as well for the same reason.
The second requirement is that of \emph{sequencing}, where there are restrictions on the ordering of the temporal
extent of an action. For instance, the actions for
handling cargo and moving the vehicle need to be ordered in causally and temporally consistent way. The action
to move the vehicle cannot temporally overlap with cargo handling activities, and clearly loading cargo at
the source must happen before unloading it at the destination.
The third and last requirement is that of \emph{concurrency}, in which some action in a plan requires some
formal property of vehicle or other entities ``states'' (a \emph{fluent}) to be true throughout its execution or to overlap with it in
some specific manner. An example of this in our cargo transportation example, follows from observing that
handling cargo will require to have other robots or human workers present at the source or destination location,
ready to load or unload the cargo.
These restrictions can be represented implicitly in terms of preconditions, effects, and invariants~\cite{fox:pddl21},
given as logical formulas over sets of fluents, or
given explicitly as temporal constraints between pairs of time-points that define time intervals with fluents
associated to them~\cite{muscettola:hsts,ghallab:ixtet,bitmonnot:fape}.

\begin{figure}[t]
    \includegraphics[width=\columnwidth]{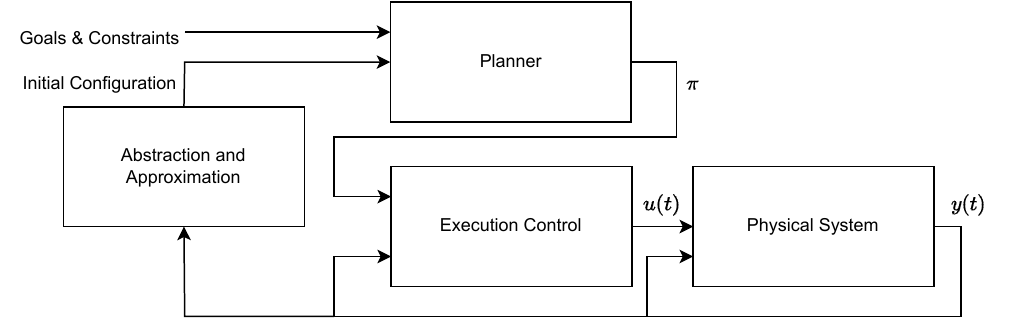}
    \caption{An architecture for (semi)-autonomous system, boxes represent physical and/or computational processes,
    arrows indicate information exchanged between these.}
    \label{fig:semi_autonomous_system_architecture}
\end{figure}

This paper proposes a formulation of temporal planning in which actions are given concrete meaning, that of
\emph{atoms} in so-called Motion Description Languages (MDL)~\cite{egersted:mdl2,egerstedt:mdl,brockett:mdl}.
Figure~\ref{fig:semi_autonomous_system_architecture} depicts an abstract architecture for semi-autonomous
systems where boxes are \emph{processes} and arrows indicate information flows, and we will refer to as
\emph{teleo-reactive}~\cite{ingrand:survey}.
The \emph{physical system} to be controlled is represented by a standard input-output model~\cite{lee:embedded_systems},
where a continuous time, multi-dimensional input signal $u(t)$ is applied on the system affecting the temporal
evolution of $y(t)$ the system output.
Autonomy in this architecture is achieved by coupling a \emph{deliberative} component, in our case a
temporal planner (TP) that comes up with plans $\pi$,
with an \emph{executive} (EC) component that defines $u(t)$ \emph{online}, scheduling the execution of MDL atoms
that actions in plans $\pi$ are mapped to.
Inputs to temporal planners are specifications, in some suitably defined formal language, of the \emph{goals}
that the plan must achieve, and \emph{constraints} on plans.
In this paper, these constraints include
the definition of the mappings between actions and MDL atoms,
the representation of the suitability, relevance, sequencing and concurrency requirements discussed earlier
in this section, and any suitable abstractions of properties of $y(t)$, or \emph{fluents}, such as its value
being inside some set for a set amount of time. Communication between the executive and deliberative components
takes place mediated by a symbolic interface, embodied in Figure~\ref{fig:semi_autonomous_system_architecture}
by a process of \emph{abstraction and approximation}, that is defined offline and determines which fluents are true
at any given point in time, and so the (abstract) initial configuration of the phyisical system to be
considered by the temporal planner.
We end these introductory remarks with the observation that two control strategies coexist in our architecture,
operating at two different time-scales. Namely, the deliberative level follows an \emph{open-loop control}
strategy, suitable for the pursuit of long-term goals, that sets the expectations on the outcomes to be
achieved by the input signal $u(t)$ generated by the executive levels.
On the other hand, the executive level follows a \emph{closed-loop control} strategy, and uses feedback from
the controlled system to compensate for disturbance while tracking the plan. The executive level may fail to
bring about the outcomes expected by plans. Such eventuality, covered by the control theoretic interpretation
of MDL atoms, and to be discussed later in the paper, puts a hold on the autonomous operation of the executive
layer, requiring a revision of the plan, either incremental, or from scratch.

\subsection{Related Work}
\label{Related_Work}

Our work belongs to the line of inquiry into temporal planning that decomposes the problem into
a causal reasoning task over suitably defined \emph{knowledge graphs}~\cite{blum:graphplan}, to identify what actions
are present in the plan, and a numerical
optimization task over (disjunctions of) linear constraints, that define the set of valid schedules for the chosen
actions. Early work focused on directly applying Blum and Furst planning graphs in temporal
settings~\cite{smith:tgp,long:graphplan,maris:tlp_gp}. The atemporal,
partially-ordered
plans that result from reachability analysis are then mapped onto a temporal constraint satisfaction problem (TCSP).
The resulting TCSP can then be solved off-the-shelf with standard
Optimization and Satisfiability technology for scheduling problems~\cite{hooker:10:integrated}. The flow of
information between the procedures for reachability analysis in planning graphs search and solving TCSPs is limited in these early approaches.
This issue is compounded by the inability of Blum and Furst planning graphs, designed for the STRIPS assumptions on
action structure, to capture plans with non-trivial concurrent interactions between actions~\cite{long:graphplan,cushing:07:ijcai}.

Research split into three major approaches in response:
Heuristic Search methods that hybridize approximate dynamic programming, using heuristics for
reachability analysis over \emph{relaxed planning graphs}~\cite{hoffmann:ff}, and consistency checking
of TCSPs~\cite{benton:optic,eyerich:tfd,jimenez:temporal}, Constraint Programming (CP)
algorithms~\cite{vidal:06:cpt} that integrate planning graph analysis into constraint propagation and backtracking search,
and Satisfiability Modulo Theory (SMT) approaches~\cite{shin:05:tmlpsat,rankooh:itsat,rintanen:17:ijcai,cashmore:20:jair,panjkovic:omt}
that propose their own or use off-the-shelf lazy or eager versions of the \textsc{Dpll}$(T)$ algorithm
for SMT~\cite{kroening:16:dp}. In these, the satisfiability of a Boolean theory in Conjunctive Normal Form (CNF) captures Blum and Furst reachability
analysis~\cite{kautz:satplan}, and sub-theories, usually linear arithmetic
over the rationals or the integers, are used to represent the TCSPs that follow from selecting actions in plans. All of the above provides
integrated reasoning over planning graphs and TCSPs, tackling some limitations of early decomposition approaches.

Notable works that rely on formulations of temporal planning distinct from the ones above include
 timeline-based planners~\cite{ghallab:ixtet} such as NASA's EUROPA~\cite{frank:cbaip}, or more recently, the
work led by A. Bit-Monnot~\cite{bitmonnot:fape}. All of these approaches use expressive temporal logic~\cite{allen:aij}
to represent and reason over properties of plans and rely on Optimization or SMT technology. Still, they struggle to
scale up as ``larger'' plans are required to solve instances, and their ability to capture complex concurrent interactions
is limited. Our ideas to represent complex temporal structures are inspired by this line of research while taking note
of lessons learned from existing research into action-centric PDDL 2.1 formulations of temporal planning.
Also reliant on Allen's are works is the work on the representation and analysis of plans defined
over timelines~\cite{cialdea:timelines,monica:timelines}.
We acknowledge the novelty of the timeline-based representations discussed in these papers with respect with
to earlier works discussed above, based on the same representation paradigm.
We note that our
approach formulates key concepts such as frame axioms in a very
different manner as these works, and we are focus on the problem of \emph{synthesizing} plans from
given specifications, rather than assuming that the plan is obtained from an oracle and \emph{analyzing}
whether certain properties hold throughout their execution.

\subsection{Contributions}
\label{Contributions}
This paper proposes a novel formulation of temporal planning that is aligned and consistent with the theoretical
foundations and established engineering practices~\cite{lee:embedded_systems} to design and analyze dynamical systems like those in
Figure~\ref{fig:semi_autonomous_system_architecture}. Our formulation relies on discrete timelines. This ensures that the
resulting optimization problem is a decidable one~\cite{gigante:complexity}, yet intractable in general, so the
challenge lies in proposing algorithms that scale up. We assume that the minimum duration of any time period considered in
a plan matches that of the control cycle set for the control system. We propose a novel notion of planning graphs,
encoded as a CP model, that results from the synthesis of Van~den~Briel et al.~\cite{vandenbriel:mip} Integer Programming
formulation of classical planning, and Allen’s Interval Logic. Crucially, our notion of planning graphs, or rather,
\emph{timing diagrams}~\cite{omg:timing_diagram}, overcome the
long-standing limitations~\cite{long:graphplan} of Blum \& Furst's seminal approach to express concurrent interactions
between the preconditions and effects of actions in the plan.

\section{Background}
\label{sec:Background}

\subsection{Interval Logic}
\label{subsec:Interval_Logic}
Allen's Interval Logic (IL) \cite{allen:cacm,allen:aij} is a temporal logic whose atoms follow from pairing
two classes of abstract objects pieces of time, or \emph{intervals}, and Boolean predicates.
The latter are \emph{atoms} in some decidable first-order theory~\cite{bradley:calculus} we will refer to
as \emph{domain theory}, and we denote by $T$.
Theories $T$ are given by a \emph{signature} $\Sigma$ that specifies constant, predicate and function symbols, and
their \emph{axioms}, sets of closed well-formed first-order formulas which only use symbols in $\Sigma$ and standard
logical connectives.
The domain theory signature is augmented, if necessary, with the signature of the Theory of the Integers $T_{\mathbb{Z}}$,
with \emph{standard} interpretation~\cite[Chapter 3]{bradley:calculus}.
Along with a domain theory $T$ we will also consider a \emph{finite} set of variables $V$.
All of this enables to represent and reason about the existence of time intervals with specific properties.

Time intervals are
\emph{periods} or \emph{moments} in time, given by either constants in $\Sigma$
or variables in $V$. We map the notions of time periods and moments onto those of time \emph{interval} and \emph{points},
respectively,
as formalized by~\cite{alur:96:mitl}. We start by
setting the \emph{time domain} \Time~to be a subset of the integers \ZPos,
whose smallest element is $0$.
A \emph{time interval} $I$ is a non-empty, convex subset of \Time~given by two numbers $l$,
$r$ $\in$ $\ZPos$.
We consider one kind of interval in this paper.
A \emph{half-open} interval $I$ is a set
\begin{align*}
I \coloneqq \{ t \in \Time \,:\, l \leq t < r \}
\end{align*}
and we refer to them in a compact fashion as $I=[l,r)$, where $l$ (resp. $r$) is the smallest or \emph{left} (largest or
\emph{right}) time-point in $I$, we refer to $l$ and $r$ collectively as the \emph{bounds} of $I$. We will use the notation $l_I$
and $r_I$ to refer to the points used to define an interval $I$. We denote the \emph{size} or volume of intervals $I$ by
$\vert I \vert$, and observe that $\vert I \vert = r - l$.

\begin{table}[ht]
    \centering
    \caption{The predicates of Interval Logic excluding \emph{meets}.}
    \label{tab:IL_predicates}
    \begin{tabular}{|@{}c@{}|c|@{}c@{}|c|}
        \hline
        Relation & Notation & Relation & Notation \\
        \hline
        $X$ equal to $Y$ & $X = Y$ & $X$ met-by $Y$  & $\MetBy{X}{Y}$ \\
        $X$ before $Y$ & $\Before{X}{Y}$ & $X$ after $Y$ & $\After{X}{Y}$ \\
        $X$ contains $Y$ & $\Contains{X}{Y}$ & $X$ during $Y$ & \During{X}{Y} \\
        $X$ started by $Y$ & $\StartedBy{X}{Y}$ & $X$ starts $Y$ & \Starts{X}{Y} \\
        $X$ finished by $Y$ & $\FinishedBy{X}{Y}$ & $X$ finishes $Y$ & \Finishes{X}{Y} \\
        $X$ overlapped by $Y$ & $\OverlappedBy{X}{Y}$ & $X$ overlaps $Y$ & \Overlaps{X}{Y} \\
        \hline
    \end{tabular}
\end{table}

Allen's IL is based on two basic predicates (binary relations). The first one is used to assert that an atomic formula,
or \emph{property}, $\varphi$ in \Atoms{\Sigma} \emph{holds through} time interval $I$.
We denote this by $\varphi@I$, a contraction of Allen's original notation $holds(\varphi, I)$.
We will refer to formulas $\phi@I$ where $I$ is non-singular, i.e. $\vert I \vert > 0$, as
Temporally Qualified
Assertions (TQA)~\cite{joslin:96:aaai}. The second basic predicate is a relation over interval pairs $X$ and $Y$
  that holds whenever $X$ and $Y$ are \emph{adjacent} or \emph{meeting}. We thus say that
$X$ \emph{meets} $Y$ (\Meets{X}{Y}) if both $X$ and $Y$ are closed, and there is no interval $Z$ such
that $\Meets{X}{Z}$, $\Meets{Z}{Y}$, $l_Z > r_X$ and $r_Z < l_Y$.
Twelve other predicates can be derived from $\Meets{X}{Y}$~\cite{allen:cacm}
and are shown in Table~\ref{tab:IL_predicates}.

\subsubsection{Syntax and Semantics}
\label{IL_Syntax_Semantics}
An IL sentence $\varphi$ is a quantifier-free first-order formula that includes both TQAs and atoms of predicates in
Table~\ref{tab:IL_predicates}, which we will refer to as \emph{temporal constraints},
using variables in $V$ and constants in $\Sigma$ to denote the bounds of intervals in \Time.
IL formulas are interpreted over some subset ${\cal I}$ of \textbf{half-open} intervals in \Time, and a \emph{history}
$h: \Time \times \Phi \to \{\top, \bot\}$~\cite[Chapter 3]{vvaa:plan_reasoning}, where $\Phi = \Atoms{\Sigma}$.
We say that the interpretation given by pairing set ${\cal I}$ with history $h$
satisfies an IL formula $\varphi$, denoted by ${\cal I}, h \models \varphi$, when
\begin{align*}
{\cal I}, h \models & \psi_1 \land \psi_2,\,&\mathrm{ iff }\,{\cal I}, h \models \psi_1\,\mathrm{ and }\,{\cal I}, h\,\models \psi_2 \\
{\cal I}, h \models &\psi_1 \lor \psi_2,\,&\mathrm{iff}\,{\cal I} , h\models \psi_1,\,\mathrm{or}\,{\cal I}, h \models \psi_2,\,\mathrm{or~both} \\
{\cal I}, h \models & \varphi@X & \mathrm{iff}\,h(t, \varphi) \mapsto \top,\,\mathrm{for}\,\mathrm{all}\,t \in X\\
{\cal I}, h \models &\Meets{X}{Y},\,&\mathrm{iff}\,r_X = l_Y\\
{\cal I}, h \models &\Before{X}{Y},\,&\mathrm{iff}\,r_X < l_Y \\
{\cal I}, h \models &\After{X}{Y},\,&\mathrm{iff}\,l_X > r_Y
\end{align*}
\begin{align*}
%\item ${\cal I} \models \MetBy{X}{Y}$, iff $r_Y = l_X$,
{\cal I}, h \models &\Contains{X}{Y},\,&\mathrm{iff}\,l_X < l_Y,\,r_Y < r_X\\
%\item ${\cal I} \models \During{X}{Y}$, iff  $l_X > l_Y$ and $r_X \leq r_Y$, or $l_X \geq l_Y$ and $r_X < r_Y$,
{\cal I}, h \models &\Starts{X}{Y},\,&\mathrm{iff}\,l_X = l_Y\,\mathrm{and}\,r_X < r_Y\\
%\item ${\cal I} \models \StartedBy{X}{Y}$, iff $l_X = l_Y$ and $r_Y < r_X$,
{\cal I}, h \models &\Finishes{X}{Y},\, &\mathrm{iff}\, r_X = r_Y\,\mathrm{and}\,l_X < l_Y \\
%\item ${\cal I} \models \FinishedBy{X}{Y}$, iff $r_X = r_Y$ and $l_Y < l_X$,
{\cal I}, h \models &X = Y,\,&\mathrm{iff}\,l_X = l_Y\, \mathrm{and}\,r_X= r_Y \\
{\cal I}, h \models &\Overlaps{X}{Y},\,&\mathrm{iff}\,l_X < l_Y,\, l_Y < r_X\,\mathrm{and}\,r_X < r_Y \\
    %\item ${\cal I} \models \OverlappedBy{X}{Y}$, iff $l_Y < l_X$, $r_Y > l_X$, and $r_Y < r_X$
{\cal I}, h \models & \varphi@X \bowtie \phi@Y,\,&\mathrm{iff}\,{\cal I}, h \models \varphi@{X}, {\cal I}, h \models \phi@{Y}, \\
& & {\cal I}, h \models X \bowtie Y
\end{align*}
$X$ and $Y$ above are \emph{interval variables} referring to elements of ${\cal I}$, and $\bowtie$ is one of the
temporal relations in Table~\ref{tab:IL_predicates}.
The interpretations
for $\MetBy{X}{Y}$, $\During{X}{Y}$, $\StartedBy{X}{Y}$, $\FinishedBy{X}{Y}$ and $\OverlappedBy{X}{Y}$ follow from
swapping $X$ for $Y$ in the rules given above for $\sim$, $\supset$, $\uparrow$, $\downarrow$, and $\frown$, respectively.

The language of temporal constraints allowed by the relations above is very rich, as one can compose via disjunction
any combination of the basic relations above, giving rise to $2^{13}$ composite
predicates~\cite{allen:cacm,beldicenau:allen}. Two such composite predicates are of particular interest
for reasoning about plan existence. The first one is the \emph{disjoint} relation, which we define as follows
\begin{align}
    \label{eq:Disjoint}
    \Disjoint{X}{Y} \equiv  \Before{X}{Y} \lor \Before{Y}{X} \lor \Meets{X}{Y} \lor \Meets{Y}{X}
\end{align}
and the second one being the \emph{inclusion} relation, defined as
\begin{align}
    \label{eq:Inclusion}
    \In{X}{Y} \equiv \Contains{X}{Y} \lor \Starts{X}{Y} \lor \Finishes{X}{Y}
\end{align}
We conclude the discussion of IL by introducing the formal properties that serve as the foundations of our
methods for reasoning about plan existence. Using~\eqref{eq:Inclusion} Allen introduces the following \emph{homogeneity axiom}
\begin{axiom}
Let ${\cal I}$ and $h$ be a set of intervals and a history as defined above. Whenever formula $\varphi$ holds
over an interval $X$, i.e. ${\cal I}, h \models \varphi@X$, then $\varphi$ holds during each subinterval $Y$
    \begin{align}
        \label{eq:Homogeneity}
        \varphi@X \leftrightarrow \big( \forall Y.\, X \sqsupset Y \rightarrow \varphi@Y \big)
    \end{align}
\end{axiom}
and  shows~\cite[Chapter 1]{vvaa:plan_reasoning} the Theorem below to follow
\begin{theorem}
    \label{thm:mutex}
Let ${\cal I}$ and $h$ be as above, $\varphi$ a formula in \Atoms{\Sigma}, and $\vdash$ a
proof system based on resolution. Whenever (1)
${\cal I}, h \models \varphi@X$, (2) ${\cal I}, h \models \psi@Y$ and (3) $\psi \land \neg \varphi \vdash \bot$,
are true, then  ${\cal I}, h \models \varphi@X \land \psi@Y$ if ${\cal I}, h \models \Disjoint{X}{Y}$.
\end{theorem}
\begin{remark}
    The Axiom and Theorem introduced above are crucial to define the structure of plans, understood as a
    description of both fluents, properties whose truth changes throughout the execution of a plan, and
    actions, that represent processes that drive change in the truth of fluents. Allen's principle of
    homogeneity allows us to ``decompose'' any TQA into sequences of adjacent TQAs for the same atom but
    shorter time spans. Theorem~\ref{thm:mutex} grounds the concept of ``mutual exclusion'' to  serialize
    access or changes in the state of ``resources'' required by actions.
\end{remark}

\subsection{Motion Description Languages}
\label{subsec:MDL}

\begin{figure}[t]
    \includegraphics[width=\columnwidth]{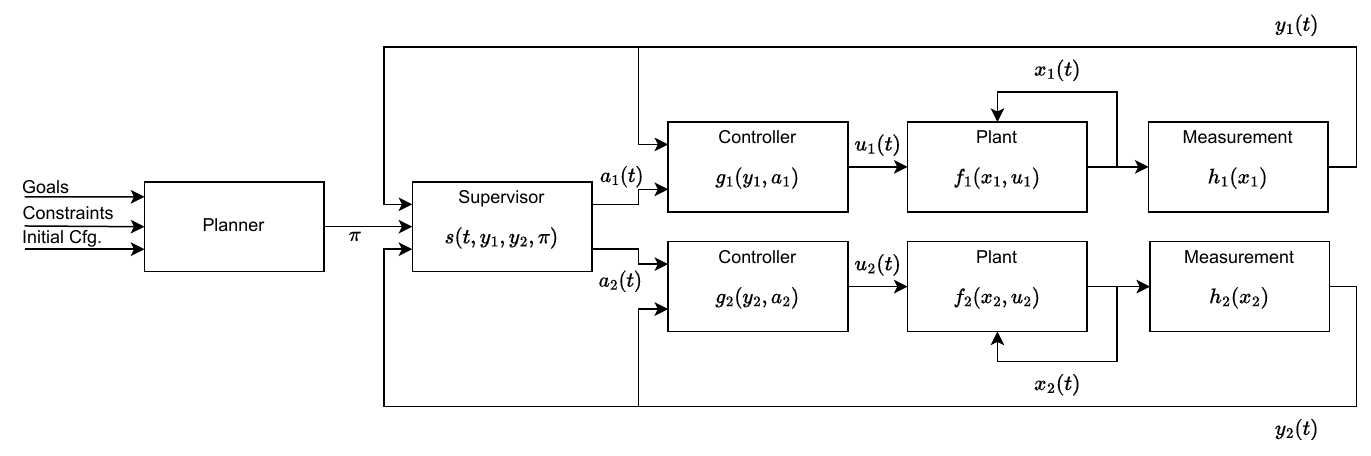}
    \caption{Refinement of Figure~\ref{fig:semi_autonomous_system_architecture} in which we identify two new
    sub-systems in the executive control (supervisor and control) and physical system models (plant and measurement).
    The diagram also makes explicit the possibility of plans defining the control for multiple autonomous systems
    that need to act in a co-ordinated manner. See text for details and discussion.}
    \label{fig:system_detailed}
\end{figure}

We now introduce the notion of Motion Description Language (MDL) following the presentation due to
Egerstedt~\cite{egersted:mdl2} for the most part, and we note below the changes we propose.
Given a finite set, or alphabet, $A$, we define a generic sequence by $\mathbf{a}$ $=$ $a_1$ $a_2$ $\ldots$ $a_m$,
which can be interpreted as partial functions from $\mathbb{N}$ to $A$ , and thus we will refer to $a_i$ by $\mathbf{a}(i)$.
The set of all such sequences is denoted by $Seq(A)$, and we denote its so-called Kleene closure by
$A^{\star}$ $=$ $Seq(A)$ $\cup$ $\{ \bm{\epsilon} \}$, where $\bm{\epsilon}$ is the empty sequence.
There is a naturally defined binary operation in this set, namely the concatenation
of sequences or strings, denoted by $\mathbf{a}_1 \cdot \mathbf{a}_2 \in A^\star$ if $\mathbf{a}_1, \mathbf{a}_2 \in A^\star$.
With respect to this operation, $A^\star$ is a \emph{monoid}, a semigroup w.r.t. concatenation with an identity element, the empty
string.
All formal languages $L$ are a subset of $A^{\star}$.

A \emph{motion alphabet} is a possibly infinite set of symbols that represent control actions that, when
applied to a specific system, define segments of motion.
A MDL is thus given by such a set of symbolic
strings that represent idealized motions, and these strings become meaningful only when the language
is defined relative to the physical system that is to be controlled.
Following~\cite{egersted:mdl2} we let the system dynamics be given by functions
\begin{align}
    \dot x &= f(x, u) \label{eq:plant}\\
    y & = h(x) \label{eq:measurement}
\end{align}
where $x$, $u$ and $y$ are \emph{functions} from $\mathbb{R}$ to some subset of $\mathbb{R}^n$.
The domains of these functions represent time, and their codomains correspond, respectively, to sets of states or
configurations $X$, inputs $U$, and outputs $Y$. We will refer to these functions as state, input and output
\emph{signals}, and to their codomains as state, input and output \emph{spaces}.
The differential equation~\eqref{eq:plant} specifies a family of mappings
$F: X^{\mathbb{R}} \times U^{\mathbb{R}} \to X^{\mathbb{R}}$, that is, of
\emph{pairs of functions} whose domains are the reals and codomains are respectively $X$ and $U$, to
a function with real-valued domain and its codomain being $X$.
Such mappings regulate how inputs change system states.
The mapping $h: X^\mathbb{R} \to Y^{\mathbb{R}}$ assigns an output value to each point $x \in X$.
Figure~\ref{fig:system_detailed} refines the notion of ``Physical System'' in
Figure~\ref{fig:semi_autonomous_system_architecture} by identifying
two subsystems\footnote{The notion of \emph{system} in Control Theory refers to any mapping between
suitably defined functions.} which we have labelled ``Plant'' and ``Measurement'', that correspond
with~\eqref{eq:plant} and~\eqref{eq:measurement}.

Egerstedt's framework is one amongst many possible approaches to decompose control strategies for complex systems
into basic building blocks, often referred to as \emph{behaviors}, \emph{modes}, or \emph{skills}~\cite{ingrand:survey}.
We will use the latter term as is the
most specific and unambiguous one
to refer to the elements of the motion alphabet $A$.
%\footnote{Egerstedt uses \emph{mode} which is associated with
% transitions in hybrid systems. We find this to be at odds with the notion of \emph{teleo-reactivity}
%articulated by Ingrand and Ghallab~\cite{ingrand:survey}.}
We say that a \emph{skill} $a \in A$ is given by a tuple $a$ $=$ $(\mu$, $\kappa$, $\xi)$ where $\mu \in U$ is the
\emph{open-loop component}, $\kappa: Y \times U \to U$ is the \emph{closed-loop component}, and
$\xi: Y \to \{0, 1\}$ is a Boolean function referred to as the motion \emph{interrupt}.
We will denote the components of a skill $a$ as $\mu_a$, $\kappa_a$ and $\xi_a$.
Given a MDL string $\mathbf{a} \in A^\star$, we denote the components of the $i$-th skill $\mathbf{a}(i)$ in
the string by $\mu_i$, $\kappa_i$ and $\xi_i$.
We also note that skills are always associated with
a \emph{single entity} in the system, or \emph{actor}. We will denote the number of actors in
the system with the letter $M$. An actor could be, for instance, a mobile robot, but also
a manipulator or a sensor on a mobile robot, that is endowed with some degree
of autonomy. A plan $\pi$ is thus an $M$-tuple of strings $(\mathbf{a}_1, \ldots, \mathbf{a}_M)$ where
$\mathbf{a}_j \in A^\star$ and $j=1,\ldots,M$, that assigns to
each actor $j$ a string. We assume that the set of plans to be a non-empty set $\Pi \subset (A^\star)^M$.
In Figure~\ref{fig:system_detailed} we have two actors, so $M=2$,
that are physically independent but their control is coupled through the assigned MDL strings,
chosen jointly to be part of a plan.

Let us consider that an actor in our system receives a string $\mathbf{a}$ of length $q$, then its
state signal $x$ satisfies the following differential equations
\begin{align*}
    \dot x & = f(x, \kappa_1(y, \mu_1)) & t_0 \leq t < T_1 \\
    & \vdots & \vdots \\
    \dot x & = f(x, \kappa_i(y, \mu_i)) & T_{i-1} \leq t < T_i \\
    & \vdots & \vdots \\
    \dot x & = f(x, \kappa_q(y, \mu_q)) & T_{q-1} \leq t < T_q
\end{align*}
where $i=1,\ldots,q$ and $T_i$ is the time at which interrupt $\xi_i$ changes from $0$ to $1$. Egerstedt
proposes a simple trigger--based hybrid system to reproduce the above
%, that interprets and operates the previous definitions as follows:
\begin{subequations}
    \begin{align}
        \label{eq:hybrid:state}
        \dot x & = f(x, \kappa_{\lfloor p \rfloor}(y, \mu_{\lfloor p \rfloor})) \\
        y & = h(x) \\
        \label{eq:hybrid:clock}
        \dot p & = \begin{cases}
               0 & \mathrm{if}\,\xi_{\lfloor p \rfloor}(t, y) = 0 \\
               \delta_t & \mathrm{if}\,\xi_{\lfloor p \rfloor}(t, y) = 1 \\
        \end{cases}
    \end{align}
\end{subequations}
where $\delta_t$ is Dirac's \emph{delta function} or unit impulse at time $t$, $x(0)$ $=$ $x_0$ is a given initial condition, and $p(0) = 1$.
In words, a new signal $p: \mathbb{R} \to \mathbb{Z}^{+}$ is introduced which is used as a \emph{clock} to track progress
in the execution of the string $\mathbf{a}$ assigned by the plan.
We note that the solution to the differential equation~\eqref{eq:hybrid:clock} will increase $p$ by one whenever
the interrupt $\xi(y)$ evaluates to $1$.
In the context of the theory of hybrid systems~\cite{tabuada:hybrid},
Egerstedt's framework is notable in that a discrete clock signal with variable rate triggers transitions between
skills, rather than considering a clock signal with constant rate.

\subsection{Supervisory Control}
\label{subsec:Supervisory_Control}

In order to address systems with more than one actor, we propose two modest changes to Egerstedt's
system model~\cite{egersted:mdl2}. First, we introduce the notion of a \emph{controller}
function $g: Y \times A \to U$ that is defined as $g(y, a) = \kappa_a(y, \mu_a)$ where $\kappa_a$
 and $\mu_a$ are the closed-loop and open-loop components of skill $a$.
Secondly, we have a explicit component that acts as an interface between the deliberative and executive levels, the
\emph{supervisor}.
The supervisor component implements~\eqref{eq:hybrid:clock}, and as depicted in Figure~\ref{fig:system_detailed}, the
supervisor is a function $s: \mathbb{R} \times Y \times \Pi \to A$ that takes as an input time, the output signals
of the $M$ actors in the system ($M=2$ in Fig.~\ref{fig:system_detailed}), and the plan chosen by the
planner to produce $M$ parallel action signals $a_{j}(t) = \mathbf{a}(p)$, identifying the skill that
is under execution at any point in time by actor $j$.

\section{Theories of Temporal Planning}
\label{section:temporal_planning_theory}

In this section we describe a framework to interpret a planning problem as that of determining the satisfiability
of an Interval Logic theory, such as the ones described in section~\ref{subsec:Interval_Logic}.
We will denote the
theory whose models describe the structure of plans and their executions by $T_{P}$.
We also observe that these theories do not exist in a vacuum, but rather are suitable abstractions of other, more
complex theories, which we denote by $T_{S}$, or \emph{system theories}.
Section~\ref{subsec:MDL} introduces the signature and some of the axioms that define such theories.
We note that more general and expressive theories can be built for complex control systems, but for the purposes of
this paper we think that Egerstedt's framework strikes the right balance between being rigorous enough to justify
our modeling choices when describing $T_{P}$, and being sufficiently flexible to account for robotic systems of
real-world interest.
Clearly, for any given system $S$ and theory $T_S$, many planning theories $T_P$ are possible, all related by
several common components and differing on choices of axioms or some elements of their signature.
Hence, what follows is best understood as a collection of axiom and symbol \emph{schemata}~\cite{bradley:calculus},
that are instantiated as it best suits to account for the particular characteristics of $T_S$.

\subsection{Assumptions}
\label{subsec:assumptions}

We start by introducing some assumptions on the structure of Egerstedt's MDL framework, which will facilitate our
discussion.
We start defining the \emph{duration} of skills $a$
\begin{definition}
    \label{def:skill_duration}
    Let $A$ be an MDL alphabet, and $\delta: A \to \mathbb{R} \cup \{ \infty \}$ be a function that maps skills $a \in A$ to
    their \emph{maximal duration}, that is the largest value that $T_i - T_{i-1}$, for $i > 1$, and $T_1 - t_0$ can take
    when $a$ appears at position $i$ (resp. $1$) of a control string $\mathbf{a} \in L_{S}$, the MDL for
    system $S$.
\end{definition}
\begin{assumption}
    \label{ass:skill_durations}
We will assume that $\delta(a)$ is always \emph{finite} and furthermore, $\delta(a) \in \mathbb{Z}^{+}$ for all
$a \in A$.
\end{assumption}
By adopting the above all the intervals $I$ will have positive and finite size $\vert I \vert$.
It does also imply that the plans for $T_P$ will be \emph{conservative} when it comes to quantify the efficiency of
plans $\pi$ with respect to their \emph{makespan}~\cite{hooker:10:integrated}, and that skill control functions
$\kappa_a$ will steer $S$ to states where $\xi_a$ is true in finite time.

Egerstedt's framework  does not interrogate the structure of skill triggers $\xi_a$.
We observe that it is commonly the case that triggers are defined by combining simpler
Boolean functions, that are used by several skills $a$.
We formalize this observation with the following
\begin{assumption}
    \label{ass:triggers}
    Let $a \in A$ be a skill in a MDL alphabet, with trigger $\xi_a$.
    We assume that $\xi_a$ is as follows
    \begin{align*}
        \xi_{a}(y) = \prod_{i=1}^{n_a} \phi_{i}(y),\; \phi_i(y) = \llbracket y \in Y_\phi \rrbracket,\;\phi_i \in \Phi_A
    \end{align*}
\end{assumption}
That is, each trigger $\xi_a$ is the product of $n_a > 0$ Boolean functions $\phi \in \Phi_A$, which we will refer to as
\emph{fluents}, and are defined to be the \emph{Iverson bracket}~\cite{knuth:notation} for the property that output
$y$ belongs to set $Y_\phi \subset Y$.
We note that the sets $Y_\phi$ need not be disjoint.

The third assumption we make is the alphabet $A$ is partitioned in the following way
\begin{assumption}
    \label{ass:precs_and_effs}
    Let $A$ an MDL alphabet.
    We assume that there exist subsets of $A$, $A_p$ and $A_e$, that satisfy the following
    conditions: (1) $A = A_p \cup A_e$, (2) $A_p \cap A_e = \emptyset$, and (3) for every
    $a \in A_p$ it holds that $\mu_a = 0$ and $\kappa_a(y, u) = 0$ for all $y \in Y$ and $u \in U$.
\end{assumption}
The above condition establishes two types of skills, those belonging to $A_p$, one whose purpose is to
monitor the evolution of
system outputs over time until  $\xi$ evaluates to $1$, and those belonging to $A_e$,
which actually provide input into the system dynamics steering them towards states in which
$\xi$ evaluates to $1$.
We will refer to the former set as \emph{precondition timers}, and we denote their elements by $p$,
while we refer to the latter set as \emph{effect delays}, which we denote with the letter $e$.

The fourth and final assumption places restrictions on the structure of control strings in $L_S$
\begin{assumption}
    \label{ass:strings}
    Let $L_S$ be MDL for system $S$.
    We assume that (1) $\bm{\epsilon} \not\in L_S$ and (2) that
    for every $\mathbf{a} \in L_S$, $\mathbf{a}(1) \in A_p$.
\end{assumption}
The above introduces the almost universal assumption in the planning literature that actions in plans are defined
in terms of a \emph{precondition} and one or more \emph{effects}, where the first skill in every control string
essentially ``waits for'' trigger $\xi_1$ to become true.
Doing so becomes very useful to account for periods of time an agent must be idle while waiting for other agents
to accomplish some task~\cite{cimatti:dcdtp}, or \emph{coasting} between two equilibrium points of its
dynamics~\cite{frazzoli:maneuver} as part of the execution of some maneuver.
We note that any MDL $L_S$ can be easily mapped into another language $L_{S}'$ that satisfies Assumption~\ref{ass:strings}.

\subsection{Signature}
\label{subsec:signature}

\begin{figure}[ht!]
    \centering
    \includegraphics[width=1.0\columnwidth]{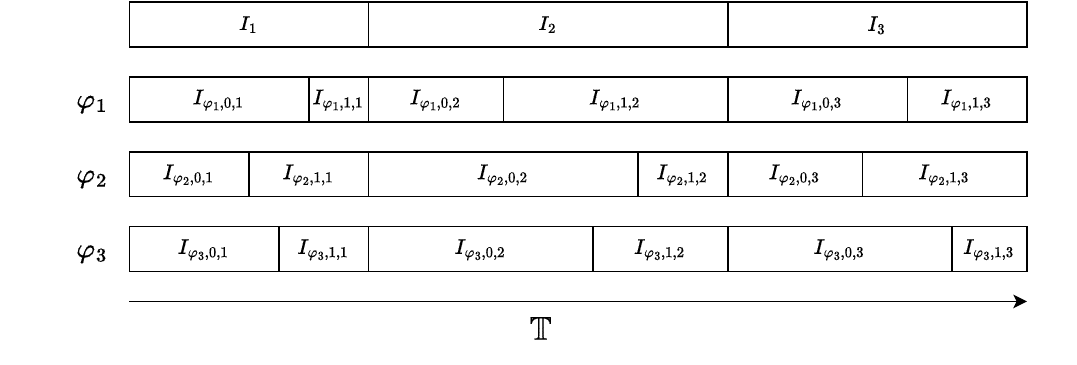}
    \caption{Global plan timeline, or \emph{dateline}, and fluent timelines. Each rectangle is an interval. The top row
    is the intervals or \emph{epochs} we use to define the temporal extent of actions. The rows below are labeled with
    the fluents that can be used with them to define TQAs. Intervals may not all be the same size but must
    satisfy other temporal constraints.}
    \label{fig:Dateline_Illustration}
\end{figure}

Temporal planning theories $T_P$ are Interval Logic theories, hence they inherit the non-logical symbols introduced in
section~\ref{subsec:Interval_Logic}.
To these we add the following predicates and constants, so TQAs (section~\ref{subsec:Interval_Logic}) can be formed.
We consider two sets of $0$-ary predicates,
${\cal F} \coloneqq \{ \varphi \,:\, \phi \in \Phi_A\}$, and
${\cal A} \coloneqq \{\alpha_j \,:\, a \in A, 1 \leq j \leq M \}$.
These predicates will be used to form \emph{fluent} and \emph{action} TQAs, describing in an abstract
manner what properties hold on the output and control components of system $S$ at every point in time.
We note that $\varphi$ and $\phi$ (resp. $\alpha_j$ and $a$) are two representations of the same object,
and thus we will use either form as it suits.
For instance, we will use always $\varphi$ and $\alpha$ in logical formulas, while we will use $\phi$ and $a$
when referring to facts about their structure, properties they may have or how they relate to other objects.

\begin{figure}[ht!]
    \centering
    \includegraphics[width=1.0\columnwidth]{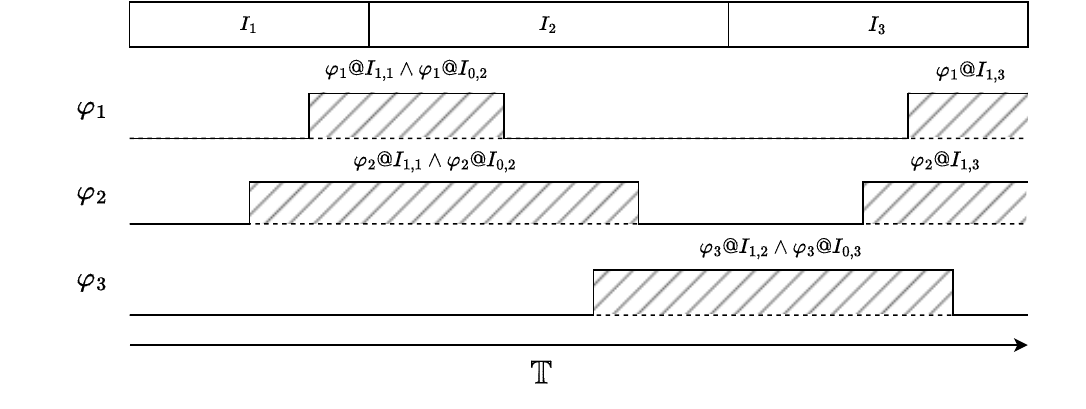}
    \caption{Truth value assignments to TQAs supported by theories $T_{P,N}$ describe \emph{timing diagrams}
    like the above. For each fluent $\varphi_i$, $i=1,2,3$, we set to true TQAs referring to the following
    intervals shown in Figure~\ref{fig:Dateline_Illustration}: $I_{\varphi_1,1,1}$, $I_{\varphi_1,0,2}$,
    $I_{\varphi_1,1,3}$, $I_{\varphi_2,1,1}$, $I_{\varphi_2,0,2}$, $I_{\varphi_2,1,3}$,
    $I_{\varphi_3,1,2}$, and $I_{\varphi_2,0,3}$.
    Hatched ``bumps'' correspond to periods of time the fluent on the left is true. Above each of
    such periods we write the corresponding IL formula.}
    \label{fig:Timing_Diagram_Example}
\end{figure}

One subtle modeling issue arises when defining the sets of intervals that need to be considered to form TQAs.
It is obvious that plans may require several instances of the same fluent and action symbols to become true
at different, non-overlapping periods of time.
In terms of MDL strings and actors, we could for instance have the same skill $a$ assigned to two different actors
$j \neq j'$ at two different moments, or alternatively, multiple non-adjacent instances of the same skill $a$ in
an actor control string $\mathbf{a}_j$.
To allow for this kind of structures in plans, from now on we consider planning theories to be parametrized both by a
positive integer $N$, and denote them as $T_{P,N}$.
We thus have the set of intervals ${\cal I}_{P,N}$, which is of finite size, and has the following structure
\begin{align}
    \label{eq:Temporal_Frame}
    {\cal I}_{P,N} \coloneqq \{ I_\prec, I_\succ \} \cup {\cal D}_{P,N} \cup {\cal S}_{P,N} \cup {\cal K}_{P,N}
\end{align}
${\cal D}_{P,N} \coloneqq \{ I_1, \ldots, I_N \}$ is a \emph{dateline}~\cite[Chapter 3]{vvaa:plan_reasoning} depicted
on the top row of Figure~\ref{fig:Dateline_Illustration}, i.e. a totally
ordered set of intervals such that $\Meets{I_{t}}{I_{t+1}}$ for $1 \leq t < N$\footnote{Technically this property
is an axiom, albeit a trivial one}.
To form action TQAs a theory $T_{P,N}$ considers the set of \emph{action intervals}
${\cal K}_{P,N}$, defined
\begin{align}
\label{eq:action_intervals}
{\cal K}_{P,N} \coloneqq \{ I_{\alpha,k} \,:\, \alpha \in {\cal A},  1\leq k < N\}
\end{align}
For fluent TQAs, $T_{P,N}$ considers the set of \emph{fluent intervals}
\begin{align}
{\cal S}_{P,N} \coloneqq \{ I_{\varphi,0,t}, I_{\varphi,1,t} \,: \, \varphi \in {\cal F}, 1 \leq t \leq N \}
\end{align}
When using the intervals above to form action and fluent TQAs we will skip the fluent (action) symbol from
the interval subindex list, e.g. we write $\varphi@I_{0,t}$ rather than $\varphi@I_{\varphi,0,t}$.
Importantly, it is only allowed to combine fluents $\varphi \in {\cal F}$ (resp. actions $\alpha_j \in {\cal A}$)
with intervals $I \in {\cal S}_{P,N} \cup \{ I_{\prec}, I_{\succ} \}$ (resp. intervals $I \in {\cal K}_{P,N}$).
We denote the set of TQAs allowed by a theory $T_{P,N}$ as $\mathsf{TQA}_{P,N}$.

We finalize this section noting that all the predicates and constants above are
\emph{uninterpreted}~\cite{bradley:calculus}.
That is, their values are to be chosen by defining an \emph{interpretation} that assigns truth values or
elements of the time domain $\mathbb{T}$ to symbols $\varphi$, $\alpha$ and $I$ in such a way that
a number of axioms and constraints, detailed below, are satisfied.
Figure~\ref{fig:Timing_Diagram_Example} illustrates how an interpretation that sets TQAs truth values
corresponds with a timing diagram~\cite{omg:timing_diagram} that tracks the evolution of binary variables
values over time.

\subsection{Connecting Planning and System Theories}
\label{subsec:system_theory_connection}
We now make the satisfiability of  $T_{P,N}$ and $T_S$ to be connected by means of the following two axioms, which
use the symbols in the signature of $T_{P,N} \cup T_S$
\begin{align}
    \label{eq:planning_and_system:1}
    \phi(t) = 1 &\leftrightarrow \big( \varphi@I \land l_I \leq t < r_I\big) \\
    \label{eq:planning_and_system:2}
    a_j(t) = a &\leftrightarrow \big(\alpha_j@I \land l_I \leq t < r_I\big)
\end{align}
By introducing the two logical equivalences above, a correspondence between the models of $T_{P,N}$ and $T_S$
is established trivially, as it is easy to see that by forcing the truth values of the left-hand sides
the rest of functions and variables in $T_S$ become all uniquely determined.
From the above it is trivial to see that there is a bijection between system output functions $y \in Y^\mathbb{R}$
and histories $h$ (see section~\ref{subsec:Interval_Logic}).
We assume that there exists at least one such function $y$ that satisfies the system theory $T_S$ dynamic
constraints discussed in sections~\ref{subsec:MDL} and~\ref{subsec:Supervisory_Control}.

The intuitive interpretation for the above is that the planner component, which is tasked with finding
satisifying assignments for $T_{P,N}$ that in turn encode plans, is able to choose what events and temporal
evolutions occur.
Plan existence thus is proof of, from the point of view of Control Theory, \emph{weak controllability}
as in the \emph{possibility} of steering the controlled system to the set of desired terminal states, rather
than certifiying the \emph{certainty} of a plan execution doing so.
Such a global liveness property is known as \emph{controllability} in Control Theory and an example analysis
for a special type of MDL languages can be found in the work of Emilio Frazzoli et al. on the \emph{Maneuver Automata}
framework~\cite{frazzoli:maneuver}.
We next describe the three types of axioms to be found on theories $T_{P,N}$.

\subsection{Type I Axioms: Actions}
\label{subsec:actions}

The first set of axioms we consider address the relation (if any) between skills $a \in A$ being executed, that is,
axiom~\eqref{eq:planning_and_system:2} holding for some actor $j$ at time $t$, and changes in the
values assigned by Boolean functions $\phi \in \Phi_A$ to system outputs $y(t) \in Y$, that is,
axiom~\eqref{eq:planning_and_system:1} truth changing between two time instants $t$ and $t'$ where $t < t'$.
These axioms are not universal, as applying to every system $S$, but depend on the suitability of the MDL alphabet
to a wide variety of control systems, such as wheeled robots or other human-designed vehicles~\cite{frazzoli:maneuver}.

Action axioms have the following general structure. For every skill $a \in A$, actor $j$ $=$ $1$,$\ldots$,$M$,
and set of intervals ${\cal I} = \{I_0 \in {\cal K}_{P,N} \} \cup \{ I_k\,: I_k \in {\cal S}_{P,N}\}_{k=1}^{n_a}$
we have that
\begin{align}
    \label{eq:axioms:actions}
    \alpha_{j}@I_{0} \rightarrow \mathsf{TC}_a({\cal I})
\end{align}
where $\mathsf{TC}_a$ is a conjunction of temporal constraints (Table~\ref{tab:IL_predicates})
between fluent TQAs $\varphi@I$ in $\mathsf{TQA}_{P,N}$ and the action TQA $\alpha_{j}@I_{0}$.
Clearly, the structure of~\eqref{eq:axioms:actions} is specific to each skill $a \in A$, as they represent
relevant properties of skills being applied, system outputs, temporal order and duration.
Additionally, in order to break undesired symmetries in satisfying assignments of $T_{P,N}$ we do
require that for every $a \in A$, actor $j$ and pair of intervals $\{ I, I'\} \subset {\cal K}_{P,N}$
\begin{align}
    \label{eq:axioms:non_overlap}
    I \frown I' \leftrightarrow \neg \alpha_{j}@I \lor \neg \alpha_{j}@I'
\end{align}
In words, the above ensures that no action TQAs assigning skill $a$ to some actor $j$ at different times
do so for periods of time that overlap.
The last set of axioms related to actions are explanatory frame axioms~\cite{shin:05:tmlpsat}, that require
some suitable action TQA to justify a fluent changing its truth value.
For every $\varphi \in {\cal F}$, and intervals ${I, I'} \subset {\cal S}_{P,N}$ we need to have that
\begin{align}
    \label{eq:axioms:frame}
    \Meets{\varphi@I}{(\neg \varphi)@I'} \rightarrow
    \bigvee_{a \uparrow \phi} \alpha_{j}@I'' \land \Contains{I''}{I} \land \Contains{I''}{I'}
\end{align}
where $a \uparrow \phi$ is a relation between skills $a \in A$ and Boolean terms $\phi \in \Phi_A$ that holds whenever
$\phi$ is a term of $\xi_a$ and is deemed that $\phi$'s value depends directly on the inputs applied by the
control function $\kappa_a$.

\subsection{Type II Axioms: Domain Constraints}
\label{subsec:domain_constraints}

The second type of axioms in planning theories $T_{P,N}$ account for theorems of the system theory $T_S$ that state
mathematical relations between sets $Y_{\phi}$ (see Assumption~\ref{ass:triggers} in section~\ref{subsec:assumptions}).
These facts about these sets are then reflected into $T_{P,N}$ as axioms, which can be used to reasoning about plans.
Of specific interest to us in this paper are  mathematical relations of the kind
\begin{align*}
    Y_{\phi_1} \cap Y_{\phi_2} = \emptyset
\end{align*}
that is, that the subsets of output space $Y$ where both Boolean functions $\phi_1$ and $\phi_2$ evaluate to $1$ are
disjoint.
From this fact it trivially follows that
\begin{align*}
    y \in Y_{\phi_1} \leftrightarrow y \not\in Y_{\phi_2}
\end{align*}
Using Theorem~\ref{thm:mutex} the following axioms follow, for fluents $\varphi_1$ and $\varphi_2$ corresponding with
$\phi_1$ and $\phi_2$, and for every pair of interval $I, I' \in {\cal S}_{P,N}$ compatible with these fluents
\begin{align}
    \label{eq:axioms:mutex}
    \varphi_{1}@I \land \varphi_{2}@{I'} \rightarrow I \mid\mid I'
\end{align}
We use the phrase \emph{domain constraints} for these axioms as they are, from our point of view, semantically equivalent
to the notion with the same name widely used in action-centric temporal planning~\cite{gerevini:constraints,shin:05:tmlpsat}
and the notion of \emph{grounding constraints} used in the literature on timeline-centric temporal
planning~\cite{frank:cbaip,cialdea:timelines}.
For convenience, we will use the notation $\phi_1 \mid\mid \phi_2$ (resp. $\varphi_1 \mid\mid \varphi_2$) for pairs of
fluents that satisfy the requirements above.

\subsection{Type III Axioms: Temporal Structures}
\label{subsec:temporal_structures}

The third set of axioms describe \emph{temporal structures} that must be satisfied by all assignments to fluent
and action TQA symbols.
In contrast with actions and domain constraints, these axioms are \emph{universal} to all planning theories $T_{P,N}$.
Time intervals in fluent TQAs must cover dateline intervals $I_t$ (see section~\ref{subsec:signature}) and correspond to
a partition each of these into two distinct, non-overlapping sub-intervals.
This property is formalised by requiring that for every fluent symbol $\varphi$ and intervals
$I_{\varphi,0,t}$, $I_{\varphi,1,t}$ in ${\cal S}_{P,N}$, the following holds
\begin{align}
    \label{eq:Dateline:FluentIntervals}
    \big(\Meets{I_{\varphi,0,t}}{I_{\varphi,1,t}}\big) \land \big(\Starts{I_{\varphi,0,t}}{I_t}\big) \land
    & \big(\Finishes{I_{\varphi,1,t}}{I_t}\big)
\end{align}
Time intervals in action TQAs must also cover the dateline intervals, but do so from the ``outside'' as in
\emph{aggregating} several of them
where each $I_{\alpha,k}$ is
\begin{align}
    \label{eq:Dateline:ActionIntervals}
    I_{\alpha,k} \coloneqq \bigcup_{l_{\alpha,k} \leq t < r_{\alpha,k}} I_{t}
\end{align}
$l_{\alpha,k}$ and $r_{\alpha,k}$ above are integer constants with domain $[1, N]$ that define the interval
$I_{\alpha,k}$ (see section~\ref{subsec:Interval_Logic}).
The size of intervals $I \in {\cal K}_{P,N}$ are subject to constraints on their size $\vert I \vert$.
Intervals $I_{\alpha,k}$ where $a \in A_p$ must be satisfy the condition that $\vert I_{\alpha,k} \vert \geq 1$,
while intervals $I_{\alpha',k}$ where $a' \in A_e$ must in turn satisfy $\vert I_{\alpha,k} \vert = \delta(a')$.
We note that the constraints above for intervals $I_{\varphi,0,t}$, $I_{\varphi,1,t}$, and $I_{\alpha,k}$
are trivial consequences of Allen's
Homogeneity Axiom~\eqref{eq:Homogeneity}.
To wit, for every IL sentence $\Phi$ using TQAs $\varphi@I$ there is a finite\footnote{$\mathbb{T}$ is a countable set.}
number of \emph{equisatisfiable} IL sentences $\Phi'$ where each occurrence of $\varphi@I$ is replaced by %the following
\begin{align}
    \label{eq:timeline_equisat}
    \bigwedge_{i=1}^{n} \varphi@I_{i} \land \bigwedge_{i=1}^{n-1} \big( \Meets{I_i}{I_{i+1}} \big)
    \land \Starts{I_1}{I} \land \Finishes{I_n}{I}
\end{align}
Importantly, intervals $I_{\prec}$ and $I_{\succ}$ are arbitrarily chosen constants, and we require the following
\begin{align}
    \label{timeline:start_and_end}
    \Meets{I_{\prec}}{I_1} \land \Meets{I_n}{I_{\succ}}
\end{align}
from dateline intervals $I_1, I_n \in {\cal D}_{P,N}$.
Associated with intervals $I_{\prec}$ and $I_{\succ}$ we have \emph{initial} and \emph{terminal} conditions
$\Phi_0, \Phi_{\star} \subset \Phi_A$, so for each $\varphi \in \Phi_0$ and $\varphi' \in \Phi_{\star}$
the following must be true
\begin{align}
    \label{eq:initial_and_terminal}
    \bigwedge_{\varphi \in \Phi_0} \varphi@I_{\prec} \land \bigwedge_{\varphi' \in \Phi_{\star}} \varphi'@I_{\succ}
\end{align}

\subsection{Plan Existence and Satisfiability}
\label{subsec:plan_existence_and_satisfiability}

With the signature and axioms of $T_{P,N}$ we can now discuss its satisfiability.
The latter is contingent on the existence of plans, objects with the following structure
\begin{definition}
    \label{def:plan}
    Let $\mathsf{TQA}_{P,N}$ be the set of TQAs for a theory $T_{P,N}$.
    A plan is a \emph{partial} function
    $\pi: \mathsf{TQA}_{P,N} \to \mathbb{B}\times \mathbb{Z}^{+}_{0} \times \mathbb{Z}^{+}_{0}$ so, given
    TQA $\psi_I$ then $\pi(\psi_I) = (b, l, r)$ where $b$ is a Boolean value that gives the truth value of
    of fluent or atom $\psi$, and $l$ and $r$ are non-negative integers that define $l_I$ and $r_I$.
\end{definition}
We say that a TQA $\psi_I$ is \emph{in the plan} if and only if $\pi(\psi_I)$ is defined.
We will denote by $\Pi_{P,N}$ the set of all objects that satisfy the Definition above.
$T_{P,N}$ is thus satisfiable if there is some $\pi \in \Pi_{P,N}$
that satisfies all axioms and constraints in $T_{P,N}$.
Intervals ${\cal D}_{P,N}$ are thus best understood as \emph{auxiliary variables} that enable us to formalize the
temporal structure axioms in section~\ref{subsec:temporal_structures}, their values determined by the values
taken by intervals in ${\cal S}_{P,N}$ and ${\cal K}_{P,N}$.

\section{Satisfiability of Temporal Planning Theories via Constraint Programming}
\label{sec:Plan_Search_CP_IP}

We now give algorithms for (1) determining the least value for the parameter $N$ so that $T_{P,N}$ is satisfiable,
and (2) to determine whether a theory $T_{P,N}$ is satisfiable, via an encoding into Constraint
Programming\cite{stuckey:constraints,hooker:10:integrated,kroening:16:dp}, or CP for short.

As observed in section~\ref{subsec:signature} in order to articulate the notion of plan, it is required first to
make a commitment on the ``size'' of the set of candidate plans to consider.
While one aspect of this size parameter is fixed by defining fluent ${\cal F}$ and action ${\cal A}$ predicate,
the number $N$ of ``moments in time`` or ``stages'' in the dateline structure of the plan~\cite{vvaa:plan_reasoning}
to consider is not given and needs to be calculated.
We search for the value of parameter $N$ in the framework of planning as \textsc{Sat}~\cite{kautz:satplan},
by considering a sequence of theories $T_{P,N_1}$, $T_{P,N_2}$, $\ldots$, $T_{P,N_k}$, $\ldots$,
defined as covered in section~\ref{section:temporal_planning_theory}.
For each theory $T_{P,N_k}$ we check if there exists some plan $\pi \in \Pi_{P,N_k}$ that satisfies all theory
axioms and constraints, and stop as soon as we found one such plan.
It is easy to see that this algorithm is complete provided that there exists a satisfiable theory $T_{P,N}$ where
$N > 0$ and finite.
Several algorithms have been proposed to accelerate this basic linear search procedure~\cite{rintanen:sat,streeter:dp},
at the expense of accepting parameters $N_k$ larger than strictly necessary, and we do not claim any specific
contribution in this regard.

To determine the satisfiability of theories $T_{P,N}$, for some suitably chosen $N$, either directly or via
a suitably defined search procedure, we give an encoding of $T_{P,N}$ into a \textsc{Csp} with variables and constraints
supported by state-of-the-art Constraint Programming solvers, such as Google \textsc{Or}-Tools~\cite{ortools}.
Importantly, the variables and constraints of the \textsc{Csp} are representations of the
uninterpreted symbols and axioms of the theory $T_{P,N}$ chosen to maximize the performance of state-of-the-art
CP solvers, rather than a direct compilation.

Let us recall Definition~\ref{def:plan} where plans are defined as partial functions $\pi$ that map TQAs in $T_{P,N}$
into triples $(b,l,r)$ of Boolean and integer values.
The fact that not all TQAs may be part of a plan $\pi$ is a non-trivial complication that we address in different
ways for fluent and action TQAs, each having associated several variables of the \textsc{Csp} that allow to choose
which TQAs are in the plan and what values they have.

For each fluent $\varphi \in {\cal F}$ and dateline index $1 \leq t \leq n$ we define Boolean variables $\varphi_{vw}^{t}$,
where $v, w \in \{0, 1\}$.
These Boolean variables determine the structure of timing diagrams (see Figure~\ref{fig:Timing_Diagram_Example})
by asserting pairs TQAs
\begin{align}
    \varphi_{00}^{t} = 1 \leftrightarrow (\neg \varphi)@I_{0t} \land (\neg \varphi)@I_{1t}
\end{align}
In words, setting $\varphi_{00}^t$ to $1$ indicates that the $\varphi$ is false for the entirety of $I_t$.
The pair of TQAs for $\varphi_3$ given in Figure~\ref{fig:Timing_Diagram_Example} would be thus asserted by
setting $\varphi_{3,01}^2 = \varphi_{3,10}^3 = 1$.
%Whenever $\varphi_{00,t} = 1$, then TQAs $\neg \varphi_{X_{\varphi,t}}$ and $\neg \varphi_{Y_{\varphi,t}}$ are present in $S$.
%If $\varphi_{01,t} = 1$, then $\neg \varphi_{X_{\varphi,t}}$ and $\varphi_{Y_{\varphi,t}}$ are in $S$, when $\varphi_{10,t} = 1$,
%then we have $\varphi_{X_{\varphi,t}}$ and $\neg \varphi_{Y_{\varphi,t}}$  in $S$, and finally
%if $\varphi_{11,t} = 1$, then both $\varphi_{X_{\varphi,t}}$ and $\varphi_{Y_{\varphi,t}}$
%are in $S$.
The approach to choose action TQAs follows directly from~\eqref{eq:Dateline:ActionIntervals}.
For each ground action $\alpha \in {\cal A}$ and $1 \leq k \leq n$ we have a Boolean variable
$u_{\alpha,k} \in \{0, 1\}$, in addition to the integer variables $l_{\alpha,k}$ and $r_{\alpha,k}$ introduced
in section~\ref{subsec:temporal_structures} to select the dateline intervals spanned by an action TQA in the plan.
Whenever $u_{\alpha,k} = 1$,
then we have one or more action TQAs $\alpha_{k}@I_{t_l}$, $\alpha_{k}@I_{t_{l}+1}$, ..., $\alpha_{k}@I_{t_{r}-1}$ in the plan,
where $l_{\alpha,k} = t_l$ and $r_{\alpha,k} = t_r$.
The following constraints ensure that $l_{\alpha,k}$ and $r_{\alpha,k}$ are set according to~\eqref{eq:Dateline:ActionIntervals}
\begin{subequations}
    \begin{align}
        \label{eq:Action:Structure:1}
        u_{\alpha, k} &\rightarrow \big( l_{\alpha,k} < r_{\alpha,k} \big)
        \land \neg u_{\alpha,k} \rightarrow \big(  l_{\alpha,k}  > r_{\alpha,k} \big) \\
        \label{eq:Action:Structure:2}
        u_{\alpha,k} &\rightarrow  u_{\alpha,k-1}
    \end{align}
\end{subequations}
Constraint~\eqref{eq:Action:Structure:1} enforces~\eqref{eq:Dateline:ActionIntervals} directly.
When no TQA $\alpha_k$ is defined by $\pi$, we set $l_{\alpha,k}$ and $r_{\alpha,k}$ in a way that they cannot be used to justify
a fluent $\varphi$ truth changing (see constraints implementing frame axioms~\eqref{def:Frame:1}--\eqref{def:Frame:2} below).
Constraint~\eqref{eq:Action:Structure:2} is redundant, and its purpose is to break symmetries.
We finalize our discussion of the variables in the \textsc{Csp} for $T_{P,N}$ by noting that recovering
``whole'' TQAs from their decompositions
via Equation~\eqref{eq:timeline_equisat} is trivial and requires linear time on $N$.
Next, we introduce
the constraints to model temporal relations between action and fluent TQAs compatible with
action schemas preconditions and constraints, as well as, objective functions, and other
temporal constraints enforcing offsets and (action) interval durations.

\subsection{Flow Constraints}
\label{Flow_Constraints}
Changes of truth are regulated with \emph{flow constraints}~\cite{vandenbriel:mip}
\begin{subequations}
    \label{def:TruthFlows}
    \begin{align}
        \label{eq:TruthFlows:Init1}
        &\varphi_{10}^1 + \varphi_{11}^1 = 1, & \; \varphi \in \Phi_0 \\
        \label{eq:TruthFlows:Init2}
        &\varphi_{01}^1 + \varphi_{00}^1 = 1, & \; \varphi \not\in \Phi_0 \\
                \label{eq:TruthFlows:Goal}
        &\sum_{v \in \{0, 1\}} \varphi_{v1}^h = 1, & \; \varphi \in \Phi_{\star}\\
        \label{eq:FlowConstraint}
        \varphi_{0w}^t + \varphi_{1w}^t &= \varphi_{w0}^{t+1} + \varphi_{w1}^{t+1}, & \; 1 \leq t < h
    \end{align}
\end{subequations}
where $w \in \{0, 1\}$. Constraint~\eqref{eq:FlowConstraint} requires continuity of truth values at the boundary
of intervals $I_t \in D_{P,N}$, disallowing $\varphi$ truth-value to change from $1$ to $0$.
Changes in truth values are only allowed within intervals $I_t$ and must be justified by an action TQA, as per
the action axioms given in $T_{P,N}$ (section~\ref{subsec:actions}).

\subsection{Precondition and Effect Constraints}
\label{subsec:PrecEff_Constraints}

Assumption~\ref{ass:precs_and_effs} partitions the set of skills $A$ into sets of \emph{precondition timers} $A_p$
and \emph{effect delays}, $A_e$, while Assumption~\ref{ass:strings} places a restriction on the structure of
the control strings in $L_S$.
While we do not place any restrictions on the structure of action axioms~\eqref{eq:axioms:actions}, clearly not
all possible combinations of temporal constraints are generally useful for control systems based on MDLs.
For any given skill $a \in A$, $TC_a$ can clearly only consist of constraints $\Contains{\varphi@I}{\alpha_j@I_{0}}$
or $\Overlaps{\alpha_j@I_{0}}{\varphi@I}$, as it can only be the case that (1) $\phi \circ y(t)$ was true before
$a$ started and continues to do so until the trigger $\xi_a$ evaluates to true, or (2) $\phi \circ y(t)$
 is true for some time $t$ after $a$ has started.
In order to facilitate the presentation of the constraints, we introduce the following auxiliary predicates
\begin{subequations}
    \label{plan_predicates}
    \begin{align}
        \ipstarts{\alpha}{k}{t} &\equiv u_{\alpha,k} \land l_{\alpha,k} = t \\
        \ipends{\alpha}{k}{t} &\equiv u_{\alpha,k} \land r_{\alpha,k} = t \\
        \ipspans{\alpha}{k}{t}{t'} &\equiv u_{\alpha,k} \land l_{\alpha,k} = t \land r_{\alpha,k} = t' \\
        \ipcontains{\alpha}{k}{t} &\equiv u_{\alpha,k} \land l_{\alpha,k} \leq t \land t < r_{\alpha,k}
    \end{align}
\end{subequations}
which assert, resp., that whenever we have a TQA for $\alpha_k$ in the plan then
(1) $\Starts{I_t}{I_{\alpha,k}}$, (2) $\Finishes{I_{t-1}}{I_{\alpha,k}}$, (3)$I_{\alpha,k}$ overlaps
both $I_t$ and $I_{t'}$, and (4) $I_{\alpha,k} \supset I_t$.

\begin{figure}[ht!]
    \centering
    \includegraphics[width=0.75\columnwidth]{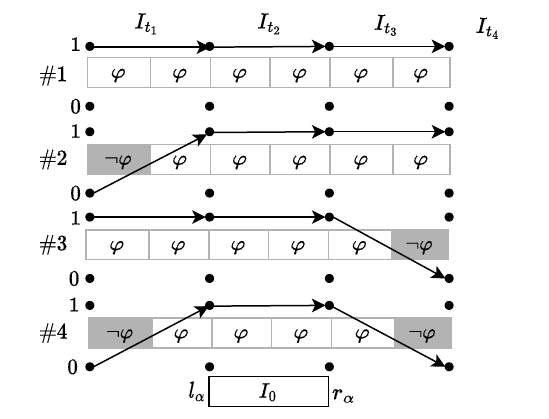}
    \caption{Illustration of the constraints enforcing temporal constraints $\varphi_I \supset \alpha_{I_0}$. From top
    to bottom, we show four possible ways to decompose $I$ as per Eq.~\ref{eq:timeline_equisat}. Epochs run in between
    vertices and selected edges in the planning graph, e.g., the arrow below $I_{t_1}$ connecting the top vertices stands for the
    decision variable $\varphi_{11}^1$ being set to $1$. Diagonal arrows correspond to a change in the truth value of $\varphi$. }
    \label{fig:Prop_Contains_Action}
\end{figure}

For each axiom like~\eqref{eq:axioms:actions} with TQA $\alpha@I$
where we have a temporal constraint $\Contains{\varphi@I}{\alpha@I_0}$,
and $2 \leq t < N$, the following constriants must be satisfied constraints
\begin{subequations}
    \begin{align}
        \label{eq:Prec:Contains:Starts}
        \ipstarts{\alpha}{k}{t} & \rightarrow \big( \varphi_{01}^{t-1} \lor \varphi_{11}^{t-1} \big)
                                         \land \big( \Meets{I_{t-1}}{I_{\alpha,k}} \big) \\
        \label{eq:Prec:Contains:Ends}
        \ipends{\alpha}{k}{t'} & \rightarrow \big(\varphi_{10}^{t'} \lor \varphi_{11}^{t'} \big)
                                         \land \big( \Meets{I_{\alpha,k}}{I_{t'}} \big) \\
        \label{eq:Prec:Contains:Spans}
        \ipspans{\alpha}{k}{t}{t'} & \rightarrow \bigwedge_{t \leq t'' < t'} \varphi_{11}^{t''}
    \end{align}
\end{subequations}
Figure~\ref{fig:Prop_Contains_Action} illustrates the four possible scenarios covered by
constraints~\eqref{eq:Prec:Contains:Starts}--\eqref{eq:Prec:Contains:Spans}, restricting when
TQAs for action $\alpha$ can be added to plans $P$. Constraint\eqref{eq:Prec:Contains:Starts}
(resp. \eqref{eq:Prec:Contains:Ends}) ensures that $I_{\alpha,k}$ starts (resp. ends) during a
time in which $\varphi$ is true. Constraint~\eqref{eq:Prec:Contains:Spans} ensures that $\varphi$ remains
true during $I_{\alpha,k}$.

For each $t$ s.t. $2 \leq t < N$ and action axiom with temporal constraint
$\Overlaps{\alpha@I_0}{\varphi@I}$ the following must be satisfied
\begin{subequations}
    \begin{align}
        \label{eq:Prec:Overlaps:Spans}
        \ipspans{\alpha}{k}{t}{t'} & \rightarrow \big( \sum_{t \leq t'' < t'} \varphi_{01}^{t''} = 1 \big) \\
        \label{eq:Prec:Overlaps:Ends}
        \ipends{\alpha}{k}{t'} & \rightarrow \big( \varphi_{01}^{t'-1} \lor \varphi_{11}^{t'-1} \big)
                                    \land \big( \Meets{I_{\alpha,k}}{I_{t'}}  \big)
    \end{align}
\end{subequations}
Constraint~\eqref{eq:Prec:Overlaps:Ends} ensures that $I_{\alpha,k}$ ends during periods that
fluent $\varphi$ is true.
Constraint~\eqref{eq:Prec:Overlaps:Spans} prescribes that there must be at most
one change in the truth value of $\varphi$, from true to false, that is allowed to take place anytime
during $I_{\alpha,k}$.

\subsection{Operational Constraints}
\label{subsec:Operational_Constraints}

Constraints~\eqref{eq:Prec:Contains:Starts}--\eqref{eq:Prec:Contains:Ends} and
\eqref{eq:Prec:Overlaps:Spans}--\eqref{eq:Prec:Overlaps:Ends} suffice to account for action axioms
necessary to account for the dynamics of control systems in the MDL framework.
Nevertheless, it is very common that one would like to consider only a subset of the set of plans that satisfy
$T_{P,N}$.
For instance, it may be the case that not all plans are convenient or desirable given some contextual information
that is not available as axioms of $T_{P,N}$ regardless of any loss of performance due to ruling out
otherwise valid output functions $y(t)$.
One example for this is when system operators wants plans to comply with given structures, like those formalized as
\emph{hierarchical task networks}~\cite{pellier:22:hddl}.
Another example is that of having two or more skills $a_1$, $a_2$, $\ldots$, $a_n$ that cannot be executed
concurrently due to being desired that plans conform to some
\emph{global resource constraint}~\cite{rintanen:17:ijcai,antoy:propagators,lepape:scheduling}.

A trivial case for skills being arranged according to some hierarhical structure are the concepts of
\emph{temporal action} present in representation frameworks such as PDDL 2.1~\cite{fox:pddl21} or
NDL~\cite{rintanen:17:ijcai}.
Temporal actions simply aggregate skills in a sequential fashion.
Let us extend the signature of $T_{P,N}$ with a finite set of uninterpreted symbols $\tau \in {\cal T}$, and consider
additional axioms
\begin{align}
    \label{eq:temporal_action}
    \tau@I \rightarrow \bigwedge_{i=1}^{n_{\tau}} \alpha_i@I_i \land \Starts{I_1}{I} \land \Finishes{I_{n_\tau}}{I} \land
    \bigwedge_{i=1}^{n_{\tau}-1} \Meets{I_i}{I_{i+1}}
\end{align}
where $I$ is an interval from ${\cal K}_{P,N}$ extended to accommodate TQAs for temporal actions $\tau$.
These constraints are directly implemented by setting trivial constraints on \textsc{Csp} variables
$u_{\alpha,k}$, $l_{\alpha,k}$ and $r_{\alpha,k}$.

Handling resource constraints on plans requires too to extend the signature of $T_{P,N}$ with uninterpreted
logical symbols $\rho \in {\cal R}$, and extending ${\cal S}_{P,N}$ accordingly.
Rather than introducing new axioms we extend action axioms, since resource constraints can be implemented with
temporal constraints $\alpha_{I_0} = \rho_{I'}$,
that approximates the notion of \emph{overall effect conditions} put forward by Cushing et al.\cite{cushing:07:ijcai}
\begin{subequations}
    \begin{align}
        \label{eq:Eff:EqualTo:Starts}
        \ipstarts{\alpha}{k}{t} & \rightarrow \rho_{01}^{t} \land \big( \Meets{I_{t-1}}{I_{\alpha,k}} \big)
                                        \land \vert I_{\rho,0,t} \vert = 1 \\
        \label{eq:Eff:EqualTo:Ends}
        \ipends{\alpha}{k}{t'} & \rightarrow \rho_{10}^{t'-1} \land \big( \Meets{I_{\alpha,k}}{I_{t'}} \big)
                                        \land \vert I_{\rho,1,t'} \vert = 1 \\
        \label{eq:Eff:EqualTo:Spans}
        \ipspans{\alpha}{k}{t}{t'} & \rightarrow \bigwedge_{t < t'' < t'-1} \rho_{11}^{t''}
    \end{align}
\end{subequations}
Namely, the TQA for resource $\rho$ is required to ``expand'' covering all the dateline intervals spanned by the action
TQA.
We also require the intervals $I_{\rho,0,t}^\varphi$ and
$I_{\rho,1,t'}^\varphi$ to span exactly $1$ time unit.
We note that to be consistent with the hybrid dynamics in~\eqref{eq:hybrid:clock}, the TQA for this effect
must start on the first control cycle after the skill $a$ starts being applied and
finishing on the control cycle \emph{before} $a$ ceases to be active.
As per the definition of MDL skills $a$, one or more functions $\phi$ will need to be observed to change
from false to true for $a$ execution to finish, hence why we allow for some minimum amount of time separation.

\subsection{Frame and Interference Constraints}
\label{Frame_Interference_Constraints}

Frame axioms ensure that no atom $\varphi$ changes  truth value without proper justification, as per
the frame axioms~\eqref{eq:axioms:frame} of the theory
\begin{subequations}
    \label{def:Frame_Axioms}
    \begin{align}
        \label{def:Frame:1}
        \varphi_{01}^t \rightarrow \bigvee_{a \uparrow \phi} \ipcontains{\alpha}{k}{t} \\
        \label{def:Frame:2}
        \varphi_{10}^t \rightarrow \bigvee_{a \downarrow \phi} \ipcontains{\alpha}{k}{t}
    \end{align}
\end{subequations}
where $a \uparrow \phi$ was defined in section~\ref{subsec:actions}, and $a \downarrow \phi$ is the relation
that holds whenever we have $a \uparrow \phi'$ and $\IntRel{\phi}{\phi'}$.
We note that when $r_{\alpha,k} = t$, we are implying that $\Meets{I_{\alpha,k}}{I_t}$, and therefore $I_t$ cannot be
contained in $I_{\alpha,k}$, since we want only to allow actions to explain those changes
that occur strictly during their intervals.

Domain constraint axioms~\ref{subsec:domain_constraints} are implemented as follows.
Let $\varphi_1\mid\mid\varphi_2$ and
$t$ be s.t. $1 \leq t \leq N$.
We need to ensure that sub-intervals $I_{\varphi_1,0,t}$,
$I_{\varphi_1,1,t}$, $I_{\varphi_2,0,t}$,  and $I_{\varphi_2,1,t}$
do not overlap \emph{when} they are referenced by a TQA in the plan, as per the definition
of $X \mid\mid Y$ in Eq.~\eqref{eq:Disjoint}
\begin{subequations}
    \begin{align}
        \label{eq:TemporalInterference:1}
        \varphi_{01}^t \rightarrow \big( \Meets{I_{\varphi_2,0,t}}{I_{\varphi_1, 1, t}}
        \lor \Before{I_{\varphi_2,0,t}}{I_{\varphi_1, 1, t}} \big) \\
        \label{eq:TemporalInterference:2}
        \varphi_{10}^t \rightarrow \big( \Meets{I_{\varphi_1,0,t}}{I_{\varphi_2,1,t}}
        \lor  \Before{I_{\varphi_1,0,t}}{I_{\varphi_2,1,t}} \big)
    \end{align}
\end{subequations}
Constraints~\eqref{eq:TemporalInterference:1} and~\eqref{eq:TemporalInterference:2} intuitively account with
the ``pushing and shoving'' amongst intervals depicted in Figure~\ref{fig:Dateline_Illustration}. We also add the
following constraints
\begin{subequations}
    \label{def:IncompatibleFlows}
    \begin{align}
        \label{eq:Interference:1}
        \varphi_{v1}^t + \phi_{w1}^t &\leq 1,\; v, w \in \{0, 1\}\\
        \label{eq:Interference:2}
        \varphi_{1v}^t + \phi_{1w}^t &\leq 1,\; v, w \in \{0, 1\}
    \end{align}
\end{subequations}
Constraints~\eqref{eq:Interference:1}--\eqref{eq:Interference:2} are meant to leverage Boolean unit propagation to discover
conflicts early due to the instance $\Int$~relation.

\subsection{Objective Functions}
\label{section:objective_functions}

Beyond finding plans for theories $T_{P,N}$ one would like also to consider the \emph{optimization} problem in
which, for a set value of $N$, we seek a plan that satisfies $T_{P,N}$ that also minimizes some given measure of
performance $f: \Pi \to \mathbb{R} \cup \{+\infty\}$, that assigns positive real values to satisfying plans
and $+\infty$ to plans that do not satisfy all axioms and constraints.
To do so we consider two classic objective functions from the literature on Optimization methods for scheduling~\cite{hooker:10:integrated}:
the so-called \emph{sum of task costs}, and \emph{makespan}.
We replace the notion of task from the literature in
scheduling with that of skill and thus assume that we have a function $c: A \to \mathbb{R}^{+}$ that assigns to
every skill $a \in A$ some positive real number.
The ``sum of skill costs`` objective function $f_{ssc}$ is thus defined
\begin{align}
    \label{eq:sum_of_costs}
    f_{ssc}(\pi) \coloneqq \sum_{a} \sum_{1 \leq k \leq N} c(a) u_{\alpha,k}
\end{align}
Formalizing the makespan objective function $f_m$ is less straightforward and requires the introduction of an auxiliary
variable $\Delta$ and auxiliary constraints
\begin{align}
    \label{eq:makespan_constraints:1}
    u_{\alpha,k} & \rightarrow \big( \Delta \geq r_{I_{\alpha,k}} \big),\; 1 \leq k \leq N, \alpha \in {\cal A} \\
    \label{eq:makespan_constraints:2}
    \Delta & \geq 0
\end{align}
with $f_m(\pi) \coloneqq \Delta$.
We also note that subsets of satisfying plans can be selected by setting \emph{deadlines} for plan completion.
This would amount to set suitable constraints on $\Delta$.

\section{Evaluation}
\label{Evaluation}

\begin{figure}[ht!]
    \centering
    \includegraphics[width=1.0\columnwidth]{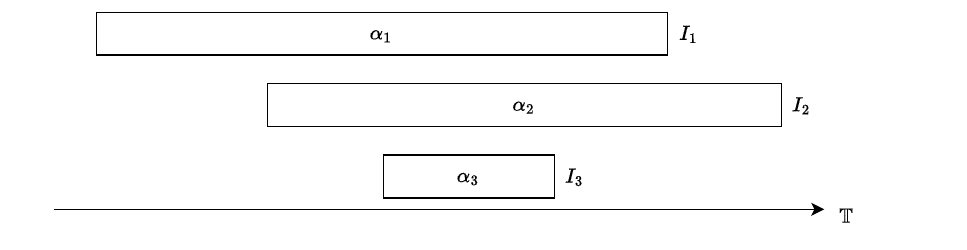}
    \caption{Action TQAs in plan for Cushing's gadget, adapted from Figure 3 in~\cite{cushing:07:ijcai}.
    The 2018 International Planning Competition benchmarks consist of instances where the plans required
    consist of increasing number of copies of the structure above, starting with $1$ and up to $18$.}
    \label{fig:Cushing_Gadget}
\end{figure}

\begin{figure}[ht!]
    \centering
    \includegraphics[width=1.0\columnwidth]{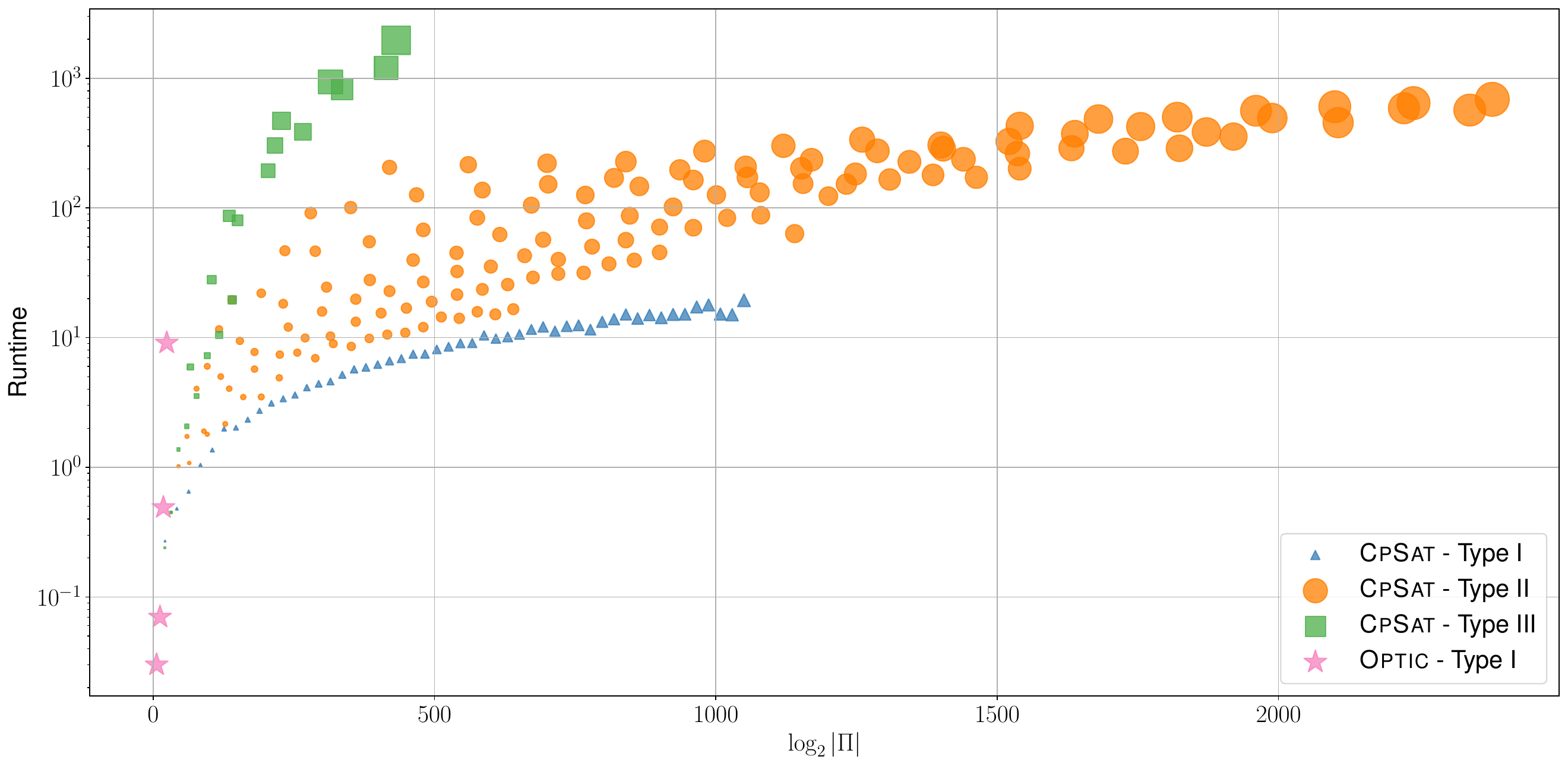}
    \caption{Run-time of our \textsc{Cp} algorithm across three types of instances. Type I instances replicate
    the structure observed in the 2018 IPC \texttt{cushing} benchmark, but with number of copies of Cushing's gadget
    going all the way to $50$.
    Type II instances require plans in which several copies of action structure in Figure~\ref{fig:Cushing_Gadget} need
    to be sequenced. Type III instances add to the previous requirement that synchronizing the executions of Cushing's
    gadgets in a hierarchical way. See text for further details and discussion.}
    \label{fig:Scalability}
\end{figure}

We have tested the previous algorithm on several benchmarks, of which we report the one we find more significant:
a generalization of the ``required concurrency'' gadget discussed by Cushing et al.~\cite{cushing:07:ijcai} to analyze
the ability of temporal planning algorithms to reason with concurrency requirements.
Cushing's gadget is crafted to only admit plans like that in Figure~\ref{fig:Cushing_Gadget}.
The gadget is requires to coordinate the execution of three skills $a_1$, $a_2$ and $a_3$, with the requirement
that the action TQAs intervals must have the configuration given in Figure~\ref{fig:Cushing_Gadget}.
We note that there are many ways to enforce such structure in plans.
In our framework, this can be achieved in two ways.
One is to do so directly by having axioms that are conditional on the presence of
any combination of action TQAs $\alpha_{i,k}@I_i$ for $i=1,2,3$ in the plan.
We note that doing so would result in having to generate $O({N}^3)$ \textsc{Csp} constraints.
A more compact modeling approach is to introduce \emph{resource} fluents (section~\ref{subsec:Operational_Constraints})
$\rho_i$ where $i=1,2$, and action axioms
\begin{align}
    \label{eq:cushing:axioms}
    \alpha_1@I_{1} &\rightarrow \rho_1@I = \alpha_1@I_1 \\
    \alpha_2@I_{2} &\rightarrow \big(\rho_2@I' = \alpha_2@I_2 \big) \land \big( \Overlaps{\rho_1@I}{\alpha_2@I_2} \big)\\
    \alpha_3@I_{3} &\rightarrow \big(\Contains{\rho_1@I}{\alpha_3@I_3}\big) \land \big(\Contains{\rho_2@I'}{\alpha_3@I_3} \big)
\end{align}
We observe that the last constraint above is not handled by many otherwise
highly performant solvers~\cite{vidal:06:cpt,eyerich:tfd,jimenez:temporal}, which ignore the requirement altogether
and thus produce invalid plans, or greatly impacts the performance of other planners~\cite{rankooh:itsat,panjkovic:omt}.
We note that Cushing's gadget can be ``scaled up'' in two obvious ways.
One is to add more ``levels'' to it, so the
``bottom'' action $a_n$ ($n=3$ in Figure~\ref{fig:Cushing_Gadget}) needs to be contained in $n-1$ intervals simultaneously.
The other way is to simply require plans to feature multiple disjoint copies of the gadget, which is the one taken
to construct the \texttt{cushing} benchmark in the temporal track of the 2018 International Planning Competition.

While Cushing's gadget is an excellent vehicle for testing the ability to reason about plans with concurrency
requirements, it ignores altogether the other key aspect of planning that we highlight in the Introduction, that
of having to sequence actions in plans.
For that we consider two additional types of structures, which we refer to as Type II and Type III.
Type II structures include Type I ones and further scale the complexity of planning ``stacking'' copies of the gadget.
That is, introducing a resource fluent $\rho_{3}^{j}$ for copies $j=1,...,m$ we introduce temporal constraints
$\alpha_{1}^{j+1}@I \rightarrow \Contains{\rho_3@I'}{\alpha_{1}^{j+1}}$.
This requires the execution of all actions in gadget $j+1$ to be contained by the time interval for the TQA of
action $\alpha_{3}^{j}$.
Type IIU structures require plans to consider $n$ copies of the gadget which need to be \emph{sequenced}.
We do so by introducing additional resource fluents $\gamma$ and temporal constraints
$\alpha_{1}^{i}@I \rightarrow \Overlaps{\alpha_{1}^1@I}{\gamma_i@I'}$ and
$\alpha_{1}^{i+1}@I \rightarrow \Contains{\alpha_{1}^{i+1}@I}{\gamma_i@I'}$, where $i$ and $i+1$ are two consecutive
copies of the gadget.

We conducted our experiments over the three types of instances above.
The Type I set considers up to $50$ copies of the gadget.
Plans for these
instances can be found for a small value of $N$, or in other words, the number of unsatisfiable theories $T_{P,k}$
with $k < N$ to consider is small.
Type II instances consider up to $20$ gadget copies, and the stacking of gadgets that appear in plans
have heights between $2$ and $8$.
Type III instances are like the Type II in terms of number of gadget copies and stacking height, but additionally
require all top level actions $\alpha_{1}$ to be totally ordered according to some arbitrary order.

Figure~\ref{fig:Scalability} plots
the run-time of our algorithm ($y$-axis) as an estimate of the size of the search space for optimal plans $P$ ($x$-axis)
grows.
We have used Google's \textsc{Cp-Sat} solver in our implementation, part of the \textsc{Or-Tools} optimization
framework.
Also plotted are the instances solved by the only PDDL 2.1 known to produce valid plans for Cushing's
gadget, Benton \cite{benton:optic} \textsc{Optic} solver.
All algorithms were allowed $1800$s and $4$ GBytes
of RAM to find a solution.
The size of data points for our \textsc{Cp-Sat} implementation depends on the number
of Boolean variables in the CP model, the largest points corresponding to CP models with hundreds of thousands of
Boolean variables.
The \textsc{IT-SAT} planner~\cite{rankooh:itsat} was not able to solve any instances, running out of memory even
with the smallest of the Type I instances.
We also tested recent planners like \textsc{Tamer}~\cite{panjkovic:omt} and \textsc{Aries}~\cite{bitmonnot:aries}.
\textsc{Tamer} solved $1$ Type I instance, and \textsc{Aries} solved $4$ Type I instances.

\section{Discussion}
\label{sec:Discussion}

Figure~\ref{fig:Scalability} clearly shows the \textsc{Cp-Sat} implementation of our algorithm to scale up for
Type I instances, and solve mid-size Type II instances in tens of seconds.
Larger Type II instances and
Type III instances do challenge our algorithm and present a motivation for further research on the
search and inference algorithms used by \textsc{Cp-Sat}, as well as on the formulation~\ref{section:temporal_planning_theory}
and encodings~\cite{sec:Plan_Search_CP_IP} that are novel contributions reported in this paper.
On the other hand, our experiments on instances where plans require actors to execute long control strings, e.g. have
non-trivial sequencing requirements, clearly show the limitations of the plan search strategy outlined at the
beginning of section~\ref{sec:Plan_Search_CP_IP}.

We look forward to drawing on the extensive work in
for Planning as SAT~\cite{rintanen:sat}, and existing lower-bounding, constraint propagation, and
decomposition techniques in Optimization~\cite{antoy:propagators,hooker:10:integrated}.
We think that our experimental results are a strong signal
highlighting the need to come up with new algorithmic ideas.
These will be needed to extend existing \textsc{Cdcl}-based solvers from the \emph{inside},
adding specific decision procedures for suitably identified subproblems, rather than just from the \emph{outside},
via encodings and restart heuristics.
Alternative solver architectures, such as those based on heuristic search~\cite{benton:optic,eyerich:tfd}, seem to us
ill-equipped to be
the framework integrating the mix of algorithms required to reason effectively over the three aspects (task assignment,
action sequencing and concurrency) of planning identified in the Introduction, while also optimizing an
objective function.
Nevertheless, heuristic search remains a key algorithmic framework for reasoning efficiently over long sequences of
actions and drive branching over decision variables, so we are sure it will remain a relevant technique delivering
critical components for next generation temporal planning algorithms.

A key motivation of this work is that of developing a rigorous yet workable analytic and computational framework
that is useful for both finding plans and controlling their execution.
For that, we choose a very specific framework from the literature in Control Theory and Robotics.
Many such frameworks have been proposed since the start of the century, with greater expressive power than the
one we have chosen in this paper.
We observe that expressive power is not the only concern one should consider when developing frameworks for research
into autonomous systems: analytical tractability (e.g. decidability of elementary properties) and computational
viability (e.g. existence of general, scalable exact and approximate algorithms to find models of such properties) are in our
opinion more important and useful for developing practical engineering applications.

We would like to end this paper acknowledging the many seminal contributions this work has been constructed upon.
Their existence and success, no matter how belatedly the latter comes to pass, are both a source of inspiration and  a challenge.

%\appendix
%\section*{Appendix A. This is a STUB.}
%[section ommitted]

\bibliography{crossref,tp_timing_diagrams}
\bibliographystyle{ieeetran}

\end{document}